# The true colours of white light: hands-on optical spectro*metry*

Robert Fischer
Optics and Materials Group, Universidade Federal de Alagoas;
Caixa Postal 2051, 57061-970, Maceió, AL, Brazil

**Abstract**
*Although the observation of optical spectra is common practice in physics classes, students are usually limited to a passive, qualitative observation of nice colours. This article discusses a diffraction-based spectrometer that allows students to take quantitative measurements of spectral bands. Students can build it within minutes from generic low-cost materials. The spectrometer's simple, didactic design allows students to fully comprehend the underlying physical concepts and to engage in a discussion on measurement errors and uncertainties.*

Optical spectrometry is a very important technique in various fields of science and technology, which is reflected by the frequent references to it in physics, chemistry, and occasionally biology textbooks. Since optical spectra are very appealing to the eye, spectroscopy is commonly used in hands-on experiments that aim to excite students for science. However, most of these hands-on experiments are using spectro*scopes* [1,2] or spectro*graphs* [3,4] rather than spectro*meter*, meaning that the student might be able to see (and record) beautiful colours, but will not obtain any quantitative results. In order to overcome this problem, chemistry outreach programs in the UK and Slovenia allow teachers to borrow an electronic spectrometer for experimenting in class [5,6]. While these programs focus on the use of spectrometry in chemistry and biology, the underlying physical effects – though mentioned in accompanying literature for the instructor – often remain hidden in an automated 'black box' attached to a computer.

Yet, with just a little bit of extra effort, it is possible to let students build their own spectro*meter* and make surprisingly accurate measurements. Students thus can gain a far deeper understanding about spectroscopy and foster their research skills by crafting and utilizing a device that measures quantities at a nanometre scale.

This article describes a simple type of spectrometer for hands-on experiments made only of a diffraction grating [7], translucent scotch tape and a printout. It has been developed to make students apply the knowledge they gained from Young's double-slit experiment [8] in a practical experience with diffraction on a grating. More specifically, this experiment challenges students to find the formula needed to calculate the central wavelength of a spectral band by employing the Huygens-Fresnel principle known from the double-slit case. Making this connection involves a repetition of the basic concepts of diffraction and helps students to see the relevance of diffraction in today's technology.

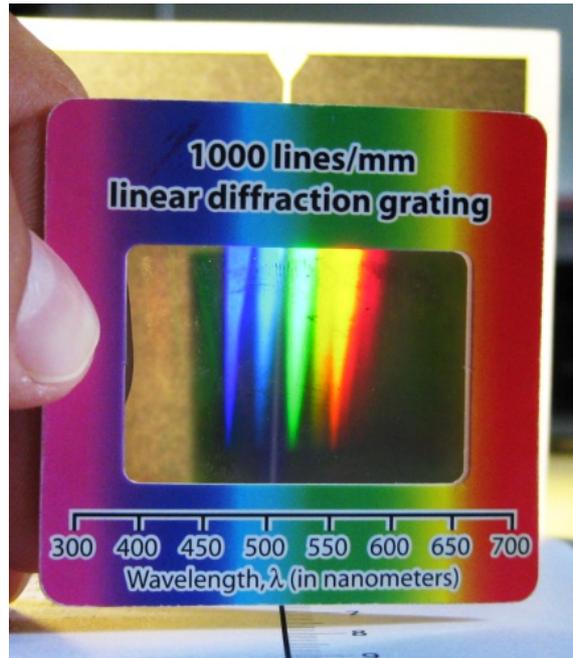

*Figure 1: View through the grating during a measurement*

**Layout and use of the spectrometer**

The layout of the printout can be generated in a few minutes on a computer using any common graphics software (e.g. OpenOffice Draw). A black box with a height of about 9 to 10 cm is drawn at the top of the page, covering the full width of the page (see Figure 2). In the middle of the box, a narrow triangle is placed with its tip pointing down. About $a$ = 4.5 to 5 cm to the right (or left, if used by a left-handed person) of the triangle is a clearly visible vertical white line. Under the box, this white line is extended with a scale down to the bottom of the page, showing the distance to the bottom of the black box e.g. in centimetres.

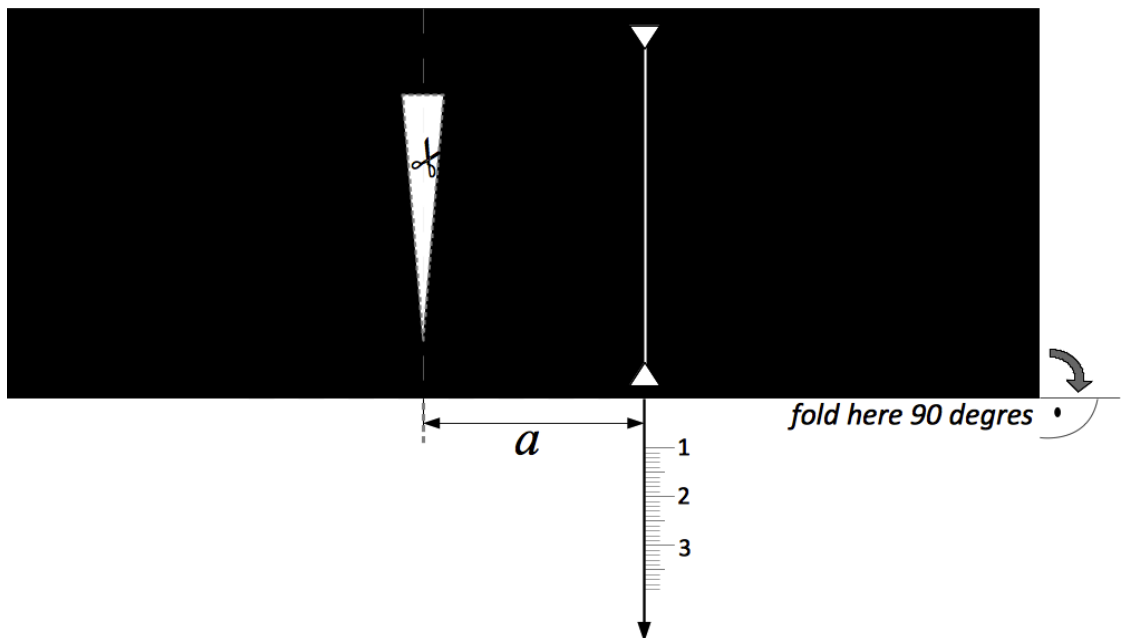

*Figure 2: Layout of printout for the spectrometer (not in scale)*

Since the measurement procedure is mastered best in teamwork, it is recommended to hand out one printout per group of two to three students. The students would need to cut out the triangle and fold the black box exactly upright along its bottom. The triangular cut-out is then covered by translucent tape, which acts as diffuser and increases significantly the visibility of spectral bands during the measurement.

Compact fluorescent light bulbs (5 to10 Watt) have proven to be very suitable for testing this student-made spectrometer. While incandescent light bulbs feature a continuous spectrum, such fluorescent light bulbs (and tubes) show clearly distinct spectral bands. Furthermore, they do not get as hot as incandescent light sources, which is an important safety issue. Because students encounter them every day, they can relate this experiment well to their personal life. Many students are surprised to see that the white light emitted by these lamps is actually a mixture of a few spectral bands, originating from phosphors on the tube's wall. Alternatively, white LEDs light sources are interesting research objects (see figure 3), and especially the screens of mobile phones have proven popular with students.

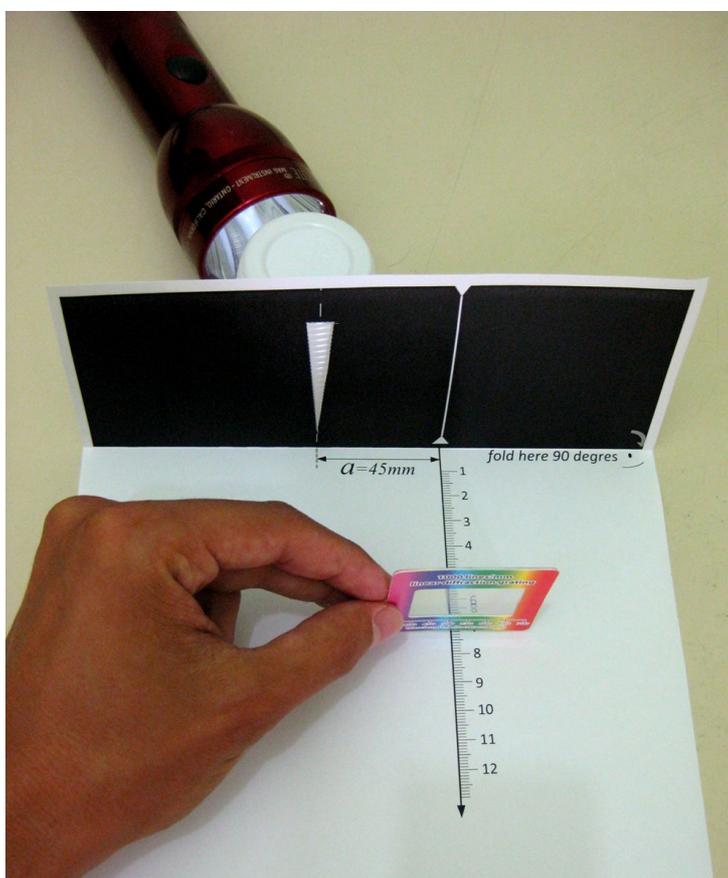

*Figure 3: The spectrometer in use, here with a LED torch and a plastic cup as diffuser.*

To measure the central wavelength of these spectral bands, the students would need to install the lamp behind the triangular cut-out. A diffraction grating with 1000 lines per mm should be placed a bit above the 7 cm mark of the scale, and held exactly parallel to the upright standing black box. Seen through the grating along the scale towards the white line, each spectral band appears as a coloured triangle against a black background (see Figure 1). By moving the grating back and forth along the scale, students will see these coloured triangles shifting to the left and to the right. To determine the central wavelength of any of the observed spectral bands, the grating should be shifted along the scale until the tip of the respective coloured triangle is situated exactly on top of the white line. The wavelength can then be calculated from the value the students read on the scale on the printout.

**Physical background**

As mentioned above, this spectrometer was designed to make students apply the knowledge gained with Young's double slit experiment to the field of spectroscopy. In the case of diffraction on a double slit, the Huygens-Fresnel principle together with geometrical arguments can guide students to find the relation between
- the distance between the first and zero diffraction order $a$, to
- the wavelength of the light $\lambda$,
- the slit size $d$, and
- the distance between the slit at the first diffraction order $b$.

Classroom tests have shown that students manage rapidly to formulate the equation $a/b = \lambda/d$, if they are provided with a clear drawing and are familiar with the concept of similar triangles. Later, in a class on diffraction gratings, the same geometrical argumentation based on similar triangles can thus be extended for many slits and applied to calculate the wavelength $\lambda$ of diffracted light (Figure 4).

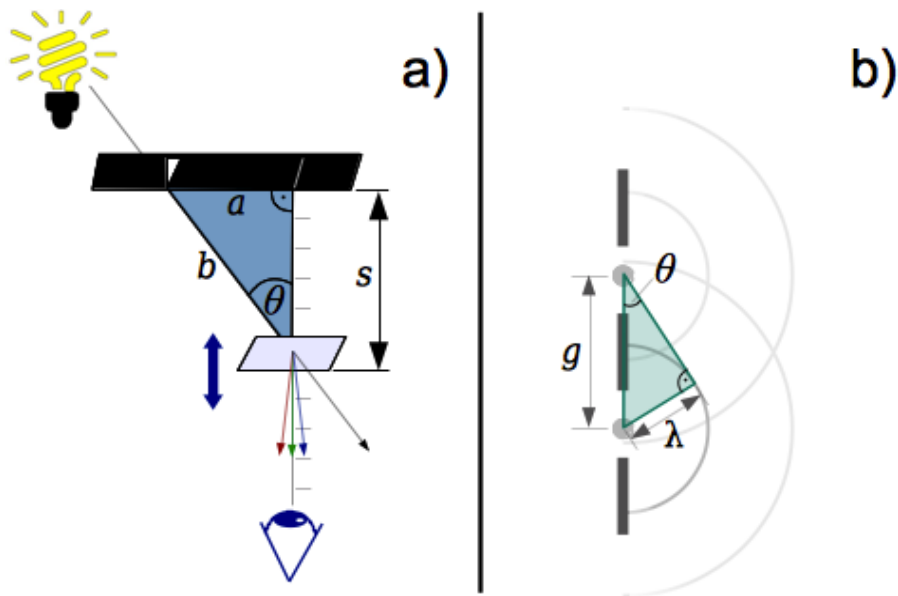

*Figure 4: a) Geometry of spectrometer; b) diffraction on grating according to Huygens-Fresnel (assuming an angle of incident to be normal to the grating)*

This geometric argumentation may seem to neglect that in the given configuration, the incident of the white light is not normal to the grating (Figure 3(a)). However, a simple thought experiment helps to see that the blue and the green triangle in Figure 3 are still similar: If the observer and the light source would swap their positions so that the angle of incident was along the normal, the observer would still see the same wavelength diffracted from the grating – yet through the triangular cut-out (without the scotch tape).

Furthermore, it can be shown that the results are actually exact within the limitations of the diffraction model commonly used at secondary schools: The grating equation for the maxima in the diffraction pattern for light under the angle of incident $\theta$ is

$$\sin(\alpha_m) = \sin(\theta) + m\frac{\lambda}{g}$$

where $g$ is the grating period, $\lambda$ the wavelength of the light, $m$ the number of the diffraction order, and $\alpha_m$ the angle of the $m^{th}$ diffraction order to the normal of the grating. In the case of the discussed spectrometer, an observer looks at the minus first diffraction order ($m = -1$) along the normal of the grating, so that $\sin(\alpha_{-1}) = 0$. Thus the expression can be simplified to

$$\frac{\lambda}{g} = \sin(\theta) = \frac{a}{b}$$

; a result which students can find by applying the Huygens-Fresnel principle along with the concept of similar triangles.

Instead of measuring $b$ directly, it is more convenient to determine first the value for $a$, and to read $s$ directly from the scale on the printout. The distance $b$ is found by applying the Pythagorean theorem, so that the central wavelength of a spectral band can be calculated as follows:

$$\lambda = \frac{ga}{\sqrt{a^2 + s^2}}$$

**Classroom experience**

The proposed spectrometer has been tested in formal and informal teaching situations with 16 to 18 years old students. In the informal setting, the building of the spectrometer and measuring of 4 spectral bands of a compact fluorescent light bulb were part of a competition. Although students were given only 15 min to complete the task (the formula to calculate the central wavelength was printed on the spectrometer sheet), the majority of groups managed to get good measurement results with at times astonishing accuracy: The root-mean square error over the 4 spectral bands of the winning group was less than 13 nm compared to the values measured by a CCD based spectrometer commonly used in industry.

In a formal classroom setting, the use of the spectrometer initiated a productive discussion on random and systematic measurement errors and measurement uncertainties. Since students build their own device, they tend to find sources of systematic errors more easily (and eagerly). For instance, if the black area of the printout is not folded upright exactly along the bottom of the black box, the value read on the scale will show an offset compared to the true value, resulting in a notable error for the calculated wavelength. The discussion and comparison of the group results also helped the students to gain a better understanding of what factors may limit the accuracy of their measurements. Besides illustrating diffraction with an eye-appealing application, the proposed spectrometer thus helps to foster crucial research and engineering skills.

**Conclusion**

Although built in a few minutes with of very simple means, the spectrometer discussed in this article allows students to make measurements with impressive accuracy. The layout of the spectrometer has been specifically designed to make it easy for students to craft their own measurement devices and to find on their own the formula needed to calculate the central wavelength of the observed spectral bands. In contrast to merely qualitative observations, students thus gain a deeper understanding of the underlying physical concepts, and see first-hand how they are applied in today's technology. Furthermore, the differences in the measurement results of different groups can initiate a productive discussion on measurement errors and uncertainties.

Besides the importance of optical spectrometry in many fields of science and technology, one should not forget that optical spectra are mainly one thing: beautiful. Their eye-catching charm seldom fails to engage students and teachers alike, which is one of the main ingredients for successful learning. The spectrometer in this article has therefore been developed to help teachers in conveying this fascination in the classroom and to add some fun for both, students and teacher.

**Acknowledgement**
The author wishes to thank Erik Stijns for helpful discussions.